\documentclass[12pt, preprint]{aastex}

\shorttitle{Effective Temperature of VY CMa}
\shortauthors{Massey et al}

\begin{document}

\title{Bringing VY Canis Majoris Down to Size: 
An Improved Determination of Its Effective Temperature}

\author{Philip Massey\altaffilmark{1,2}, Emily M. Levesque\altaffilmark{2,3}, Bertrand Plez\altaffilmark{4}}

\altaffiltext{1}{Lowell Observatory, 1400 W Mars Hill Road, Flagstaff,
AZ 86001; Phil.Massey@lowell.edu}
\altaffiltext{2} {Visiting Astronomer, Cerro Tololo Inter-American
Observatory, National Optical Astronomy Observatory, which is
operated by AURA, Inc., under cooperative agreement with the National
Science Foundation (NSF).}
\altaffiltext{3}{Massachusetts Institute of Technology, 77 Massachusetts Avenue,
Cambridge, MA 02139; emsque@mit.edu}
\altaffiltext{4} {GRAAL CNRS UMR5024, Universit\'{e} de Montpellier II, 34095 Montpellier Cedex 05, France;
Bertrand.Plez@graal.univ-montp2.fr}

\begin{abstract}

The star VY CMa is a late-type M supergiant with many peculiarities, mostly related
to the intense circumstellar environment due to the star's high mass-loss rate.
Claims have been made that would suggest this star is considerably more 
luminous ($L\sim 5 \times 10^5 L_\odot$)
and larger ($R\sim 2800 R_\odot$) than other Galactic red supergiants (RSGs).
Indeed, such a location in the H-R diagram would be well in the  ``Hayashi forbidden zone" where stars cannot be in hydrostatic equilibrium.  
These extraordinary properties, however, rest upon an assumed effective temperature
of 2800-3000 K, far cooler than recent work have shown RSGs to be.
To obtain a better estimate, we fit newly obtained spectrophotometry in the optical and NIR with the same MARCS models used for our recent determination of the physical
properties of other RSGs; we also use $V-K$ and $V-J$ from the literature to derive an
effective temperatures. 
We find that the star likely has a temperature of 3650 K,
a luminosity $L\sim 6 \times 10^4 L_\odot$, and a radius of $\sim 600R_\odot$
These values are consistent with
VY CMa being an ordinary evolved $15 M_\odot$ RSG, and agree well with the
Geneva evolutionary tracks.  We find that the circumstellar dust region has
a temperature of 760~K, and an effective radius $\sim$130 AU, if spherical
geometry is assumed for the latter. What causes this star to have such a high mass-loss,
and large variations in brightness (but with little change in color), remains
a mystery at present, although we speculate that perhaps this star (and
NML Cyg) are simply normal RSGs caught during an unusually unstable time.

\end{abstract}


\section{Introduction}
\label{Sec-intro}

VY CMa is a late-type M supergiant that is remarkable in many ways.
It has a large IR excess, making it one of the brightest objects
in the sky at 5-20$\mu$m, indicative of a dust shell (or disk)
heated by the star (Herbig 1970a).  The inferred mass-loss rate is
huge for a red supergiant (RSG), about $2\times 10^{-4} M_\odot$
yr$^{-1}$ (Danchi et al.\ 1994).   Like other late-type mass-losing stars,
it has strong molecular maser emission; 
in fact, it was one of the first radio masers discovered 
(Wilson \& Barrett 1967;  Eliasson \& Bartlett 1969).
The star is embedded in an
asymmetric dust reflection nebula which extends 8-10" from the star,
which is highly structured (Monnier et al.\ 1999; Smith et al.\
2001; Humphreys et al.\ 2005).  This nebula is so bright that it was discovered in 1917 with an
18-cm telescope, but was somehow missed by earlier observers with
much better equipment, leading to the speculation that the nebula
is only 100 years old (Herbig 1972, Worley 1972).  
The photometric history of the star extends back to 1801, and shows the star fading
visually by 2~mag since that time (Robinson 1971),  with typical
variations of $\pm2$~mag  (Robinson 1970, Henden 2006), but with little change in color. 

Spectroscopically,
the star is unusual in two regards. First, although careful long-term
monitoring by Wallerstein (1958, 1977; Wallerstein \& Gonzalez 2001)
has shown that the overall spectrum has stayed constant at M3-M5~I for over 40 years,
the spectrum shows emission of low-excitation lines, including Na I and K I,
as well as emission in the band heads of TiO and ScO (Wallerstein
1958; Herbig 1970b), which do change over the time scales of months.
Secondly, the
star's spectrum has been described as  ``veiled" (Herbig 1970b, Humphreys 1974), 
a term used to denote 
the general washing out of absorption features seen in the spectra of some
Mira variables (Merrill et al.\ 1962) and other luminous RSGs (Humphreys \&
Lockwood 1972).

However, our interest in VY CMa was prompted by the extraordinary physical
properties often assumed for this star 
(Le Sidaner \& Le Bertre 1996, Smith et al.\ 2001; Monnier et al.\ 2004; Humphreys
et al.\ 2005), such as a radius of 1800-3000$R_\odot$ (8.3-14 AU),
significantly larger than $\sim 1500 R_\odot$ value measured for the three largest
RSGs in a sample of 74 Galactic stars we have recently analyzed (Levesque et al.\ 2005, 
hereafter Paper I).   The corresponding luminosity ($2-5 \times 10^5 L_\odot$, or 
$M_{\rm bol}=-8.5$ to $-9.5$) is near or beyond the highest luminosities of other known 
RSGs (Paper I).  Usually such problems can be traced to an uncertainty in the distance, but
in the case of VY CMa the distance is relatively well determined, as the star is associated
with a molecular cloud at a distance of 1.5~kpc (Lada \& Reid 1978). This
distance is consistent
with the proper motions measured from H$_2$O maser features (Richards et al.\  1998).
Furthermore, the very large diameter 
appears to be supported by direct interferometry  (Monnier et al.\ 2004).

The extreme properties determined for this star
rest primarily upon the assumed effective temperature of 2800 K, a value
which comes from Le Sidaner \& Le Bertre (1996) who {\it adopted} this temperature
based upon the star's spectral type\footnote{Le Sidaner \& Le Bertre (1996) cite the effective temperature scale of
Dyck et al.\ (1974), but 2800~K is much cooler than what Dyck et al.\ (1974) adopt for an M4
star---roughly 3400~K; see their Fig.~7.}.
Were the star at 2800~K with $M_{\rm bol}\sim -9$ 
it would be considerably  cooler and more luminous than current evolutionary models 
allow, and would be located in the ``forbidden zone" to the right
of the Hayashi track\footnote{Note that the Hayashi tracks are analytically of the form
$\log T_{\rm eff} = 0.02 M_{\rm bol} + 0.2 \log M + C$ (Kippenhahn \& Weigert 1990; 
Bohm-Vitense 1992).  For RSGs, $\log M = 0.5 - 0.10 M_{\rm bol}$ (Paper I), and hence the edge of the forbidden zone must correspond roughly to a constant
temperature for massive stars.  The exact value of this temperature
depends critically upon the treatment of
convection; here we adopt $\log T_{\rm eff}=3.55$, where we have shifted the results of
Ezer \& Cameron (1967)  by $-0.05$~dex (consistent with a slight lowering of
their adopted value
of $l/H_P$ from 2) in order to give a better match to the evolutionary tracks of Meynet \& Maeder (2005). See also Henyey et al.\  (1965), and discussion in Kippenhahn \& Weigert (1990).} (Fig.~1).  Oddly, no previous authors have commented on this fact.
Such objects are not in hydrostatic equilibrium.  Although
RSGs would not be found in this region, collapsing protostars might, and 
indeed Wallerstein (1978) and Herbig (1970b) have long argued that VY CMa might be a 
recently-formed object rather than an evolved massive star, although based upon other
reasoning.

Understanding the nature of this interesting object rests upon knowing its effective
temperature, as this is also key to determining the star's bolometric luminosity.
We have recently used 
MARCS model atmospheres (Gustafsson et al.\ 1975,
Plez et al.\ 1992,  Plez 2003, Gustafsson et al.\ 2003)
to derive the physical properties
of Galactic and Magellanic Cloud RSGs.  Here we apply the same techniques to VY CMa,
using newly acquired optical/NIR spectrophotometry and existing IR photometry.  Thermal
emission from the dust apparently begins to dominate the spectral energy distribution past 
2$\mu$ (see Fig. 1 of Hyland et al.\ 1969), and thus we expect that the use of $V-K$ will
lead to too low an effective temperature.  However, scattering by the dust in 
the optical/NIR region may partially fill in spectral absorption features (i.e., veiling), and fitting the molecular bands using the MARCS models may lead to too
high an effective temperature.  The true effective temperature of the star should be between
these limits, and using $V-J$ may help determine which is closer.

\section{Determination of Physical Properties}

We observed VY CMa using the CTIO 4-m telescope and RC spectrograph (Loral CCD in
blue air Schmidt camera) in 
the red (5300\AA-9000\AA) using a 2~sec exposure on 20 Dec 2005, and in
the blue  (3400\AA-6200\AA) using a 30~sec exposure on 21 Dec 2005.  For both setups
we used grating KPGL2 (316 l mm$^{-1}$).  The dispersion was 2 \AA\ pixel$^{-1}$, and we
achieved a spectral resolution of 7.5 \AA\ (3.8 pixels) using a 225$\mu$m slit (1.5").
For the red observations we used an WG-495 filter to block out 2nd order
blue light.  For the blue observations, we used a BG-39 filter to remove any possibility of scattered
red light affecting our observations, especially in the far blue and near UV where
the stellar flux is small (see Massey et al.\ 2005; Levesque et al.\ 2006, hereafter, Paper II).  
The spectrograph was rotated to the parallactic angle for
each observation.  Multiple observations of spectrophotometric flux standards were obtained
throughout the night, and the agreement between these standards was typically 1-2\% after
a grey shift.  Observations of several ``featureless" stars allowed us to remove
the telluric absorption to first order, following Bessell (1999).
The data were flat-fielded using projector flat data obtained during the day, and
were wavelength calibrated using exposures of a He-Ne-Ar comparison arc obtained immediately
before (blue) or after (red) the exposure of VY CMa.  Finally, the reduced
blue and red spectra were
combined with a small adjustment to bring the flux levels in the overlap region
into agreement.   We show the spectrum of VY CMa in Fig.~\ref{fig:fit}a, along with
line identifications.

The spectra were fit using the same MARCS stellar atmosphere models used in Paper~I,
using a similar technique.  We compared the synthetic spectra to the observed one,
matching the line depths of the molecular bands (primarily TiO) and adjusting the
reddening of the model with a Cardelli et al.\ (1989) law ($R_V=3.1$).  A satisfactory fit was quickly obtained with a $T_{\rm eff}$=3650~K model, and $A_V=3.20$ (Fig.~\ref{fig:fit}b, red curve).
Our estimate on the uncertainties are 25~K for $T_{\rm eff}$ and $0.15$ on $A_V$.
This reddening is in good agreement with the $E(B-V)\sim1$ value suggested by
Wallerstein \& Gonzalez (2001).  We used a $\log g=0.0$ model for the fitting;
a model with $\log g=+0.5$ gives a better fit to the gentle red slopes of some of
the TiO bands in the NIR, but $\log g=0.0$ is more consistent with the star's inferred physical properties.  Use of the higher surface gravity model
would increase the reddening negligibly,  and not
change the temperature.  The discrepancy in the UV between the observed and expected
fluxes is similar to that seen around other RSGs with large amounts of
circumstellar dust, as inferred by excess reddening (Massey et al.\ 2005), and is likely due to
scattering of light from the star by the dust, possibly combined with a distribution of grain sizes
that is larger than normal.

We list the star's derived physical properties in Table~1.
For this, we
adopted $V=8.5$, a value close the star's average brightness, and similar to the
value the star had at the time we observed it spectroscopically, according to
estimates by the AAVSO (Henden 2006, see below).  With a distance of 1.5~kpc, and
our values for $A_V$, this determines the absolute visual magnitude ($M_V$).
The MARCS models provide the bolometric corrections as a function of effective
temperature (see Paper~I), allowing us to determine the star's bolometric luminosity
$L$.  The effective stellar radius $R$ then follows from $L=4\pi \sigma T_{\rm eff}^4 R^{2}$. 

Although the temperature is {\it significantly} higher
than that adopted by other workers, it is in accord with our own recent estimates of
the effective temperature scale of RSGs (Paper I), corresponding to a type somewhat
later than M2~I.  Although the star's spectrum has been variously
described as M3-M4~I (Joy 1942), M5~I (Wallerstein 1958), and M4-M5~I (Humphreys
1974), we would actually classify it as M2.5~I, based upon the TiO bands we normally 
use, i.e., $\lambda 6158$, $\lambda 6658$, and $\lambda 7054$ (see also
Jaschek \& Jaschek 1990).  However, our results are not really in conflict with 
earlier studies, which based their classifications primarily on blue (photographic)
spectrograms.  If instead we had used TiO bands further to the blue (such as 
$\lambda4761$, $\lambda 4954$, $\lambda 5167$, and $\lambda 5448$) we
would in fact have arrived at an M4~I classification (Fig.~\ref{fig:fit}a).   We did a fit
based upon these TiO bands, and found $T_{\rm eff}$=3450~K and a lower reddening ($A_V=2.00$). 
One can argue that the TiO bands in the blue are less likely to be affected by 
veiling (Humphreys 1974), but alternatively there is clearly excess flux in the blue
due to dust scattering (Massey et al.\ 2005).  In any event our fit to the TiO bands in the
blue was considerably poorer compared to the overall spectrum (Fig.~\ref{fig:fit} b, purple curve), and
so we are inclined to place more credence in the NIR fit.   We include
the derived physical parameters from the blue fit in Table~I for comparison.

We can also estimate the star's effective temperature using the star's $V-K$ color.
There are two difficulties with this.  The first is that the star is a notorious variable
(Robinson 1970, 1971).  In Fig.~\ref{fig:aavso} we show the variations in the visual
brightness of the star collected by the AAVSO over the years (Henden 2006), 
along with any published photoelectric $V$ values.   Oddly, the star has stayed remarkably constant
in color during these shifts; for instance, the star faded by 0.4~mag in $V$ during
the photoelectric monitoring by Cousins \& Lagerweij (1971), but remained essentially constant
in $B-V$ and $U-B$\footnote{Note that this conclusion differs from that of Robinson (1970),
who finds that the star reddens as it fades.}.  There are two IR measurements in the literature: 
$K=-0.62$ by Hyland et al.\ (1969) from November 1968, and $K=+0.34$ (after
transformation to the standard system of Bessel \& Brett 1988, following
Carpenter 2001) from 1999 November.  Despite the 0.9~mag difference, the
$J-H$ and $J-K$ colors are nearly identical: $J-H=1.49\pm0.10$ vs.\ $1.35\pm0.47$ and
$J-K=2.63\pm0.2$ vs.\ $2.26\pm0.50$, where the large errors correspond to the
2MASS colors, and are due to the photometry being derived from fitting the wings
of the stellar profile\footnote{See  
http://www.astro.virginia.edu/$\sim$mfs4n/2mass/allsky/bs\_allsky.html.}.  More
remarkably, if we take the $V=7.44$ magnitude reported by Hyland et al.\ (1969),
(which is presumably contemporaneous with their $K$ value) 
and the $V=8.40$ magnitude
that corresponds to the average of the AAVSO estimates near the time of the 2MASS observation, 
we find fortuitously identical colors,
$V-K=8.06$.  We can deredden this using $A_V=3.20$ and $A_K=0.35$, where we
have assumed that $A_K=0.11A_V$, following Schlegel et al.\ (1998), and obtain
$(V-K)_0=5.21$.  Using our calibration from Paper~I, this corresponds to an effective
temperature of $T_{\rm eff}$=3475~K.  An uncertainty of 0.2~mag in the color (not
unreasonable) would correspond to an error in the temperature of 35~K.  We include
in Table~1 the physical parameters derived from this value.

The second difficulty with using $V-K$ to estimate the effective temperature of the star is that this
will provide only a  lower limit,  given that thermal emission significantly contaminates the
$K$-band (Hyland et al.\ 1969) photometry.  A better estimate then might come from the
$V-J$ colors.  For both the Hyland et al.\ (1969) and 2MASS data sets we obtain
$V-J=5.43$ and $V-J=5.45$, respectively.  If we deredden these by $A_V=3.20$
and $A_J=0.28 A_V=0.90$ (again, following Schlegel et al.\ 1998), 
we obtain $(V-J)_0=3.14$.  Using the MARCS models
for Galactic metallicity, we find that we expect
$$T_{\rm eff}=7260.0-2073.74(V-J)_0+371.600(V-J)_0^2-22.802(V-J)_0^3$$
for $\log g=0.0$ and $4300\ge T_{\rm eff} \ge 3000$.  Thus, for VY CMa, the
$V-J$ colors imply a temperature of 3705~K.  An uncertainty of 0.2 mag corresponds
to an error of 90 K.  This is in good agreement with the 3650~K temperature we obtained
from fitting the red/NIR TiO bands.

How reliable are the temperatures derived from the MARCS models?  In Paper I
we showed that the temperatures and luminosities derived using these models brought
Galactic RSGs into accordance with the predictions of modern stellar evolutionary theory; in
Paper II we showed that this was also true
at the lower metallicities of the LMC and SMC.  We consider it unlikely that this agreement
is coincidental, and instead believe this finding provides a powerful endorsement of each.
Further, we have now shown
(Paper~II) that the MARCS models are self-consistent in that the temperatures derived
from broad-band $(V-R)_0$ are similar (to 30 K) with those derived from fitting
the TiO lines.   We do find (Paper II) that the broad-band $(V-K)_0$ colors of the models
yield effective temperatures that are $\sim 100$K warmer than those derived by the
other means.  We attribute this to the limitations of static 1D models, as spectra
of RSGs in the optical and IR may reflect different atmospheric conditions due to the
large surface granulation present in these stars.  

\section{Discussion}

We have brought the same techniques to bear on VY CMa as we have on other
RSGs.  Contrary to previous claims, we find that its effective temperature is 
typical of RSGs (3450-3700 K) and that its luminosity, while high 
($M_{\rm bol} = -7.0$ to $-8.0$, or $L=5\times 10^4$ to $1.3\times 10^5 L_\odot$) is not extraordinary.
Our preferred value comes 
from fitting the red/NIR TiO bands, which yield $T_{\rm eff}$=3650~K
and $M_{\rm bol}=-7.2$ ($L=6.0 \times 10^4 L_\odot$); this is in good agreement
with that found from the $V-J$ photometry.

Monnier et al.\ (2004) used 2-$\mu$m
interferometry to obtain a diameter of VY CMa of 18.7 mas; at a distance
of 1.5~kpc, this corresponds to a radius of 3000$R_\odot$, in seeming contradiction to the results
obtained here.  Adopting this large size leads to a temperature of 2700~K, in 
(un)fortuitous agreement with the value proposed by Le Sidaner \& Le Bertre (1996).
J. Monnier (private communication) was kind enough to offer two possible explanations.
First, their analysis assumed that the dust shell did not have significant structure on the scale of
the stellar diameter; if that were not the case, then the angular measurement might have included
part of the dust shell.  Secondly, it could be that VY CMa has a molecular water layer around the
star that emits at 2$\mu m$, similar to what is seen around some cool AGB stars.  This would cause
the continuum diameter to be larger than the photosphere. Higher resolution studies are now 
possible, and are being planned, to resolve this possible discrepancy. 

Our improved estimate of the temperature of the star has little impact on the inferred properties of the
circumstellar dust shell.   If we adopt $T_{\rm eff}$=3650~K for the star, and assume that the
$J$-band photometry is uncontaminated by the thermal emission (as is implied by the good
agreement with the $V-J$ colors and that found by fitting the red/NIR TiO bands), then we
can compute both the temperature $T_{\rm cs}$ and effective area $A_{\rm cs}$ of the circumstellar
material by comparing the dereddened IR colors with those of the stellar model.    
The dereddened
2MASS observations yield $J_0=2.07$, $H_0=1.03$, and $K_0=-0.02$.  The colors
corresponding to a MARCS $T_{\rm eff}$=3650~K, $\log g=0.0$ model are $(J-H)_{\rm mod}=0.88$,
$(J-K)_{\rm mod}=1.11$.   We assume that the observed $J_0$ is just the same as that of the model
plus a normalization constant: $J_0=J_{\rm mod} +C$, but that the $H_0$ and $K_0$ contain a flux
component due to the circumstellar shell ($F_{\rm cs}$): $H_0=-2.5\log(F_{H*} + 
A_{\rm cs} F_{H{\rm cs}})$
(with a similar equation for $K_0$), 
where $A_{\rm cs}$ is the emitting area of the circumstellar shell compared to that of the star,
and $F_{H*}$ is
the flux from the star in $H$, namely $10^{-(H_{\rm mod}+C) /2.5}$.  Then it follows that
$A_{\rm cs} F_{H{\rm cs}}/F_{H*}=10^{(H_{\rm mod}+C-H_0)/2.5}-1$.  
We find that $A_{\rm cs} F_{H_{\rm CS}}/F_{H*}=0.15$
and $A_{\rm cs} F_{K{\rm cs}}/F_{K*}=1.45$.  
The ratio $F_{K{\rm cs}}/F_{H{\rm cs}}=9.6F_{K*}/F_{H*}=11.9$, which corresponds to a
black-body temperature of 760~K.   This is close to the 850~K temperature 
deduced from 
mid-IR photometry by Le Sidaner \& Le Bertre (1996), given the uncertainties in $JHK$,
and the small leverage we have here compared to the mid-IR.
It is also consistent with the temperature range estimated by Wallerstein (1958)
for the low-excitation emission lines, and the  600~K excitation temperature estimated for
the ScO emission by Herbig (1974). 
We can go one step further, however, and compute the area based
upon the calculated fluxes per unit area of the MARCS model and a 
760~K blackbody in the $H$ and $K$ bands.
We find $A_{\rm cs}=2130$ from $H$, and $A_{\rm cs}=2155$ from $K$.  Thus, the radius of the emitting region must be about 46 times
that of the stellar radius, if it is a spherical surface. If the stellar radius is $\sim 600 R_\odot$, this circumstellar material
would have an effective radius $\sim$130 AU.

We find that the location of VY CMa in the HRD is consistent with that of an
evolved $15 M_\odot$ star.  Why then does VY CMa have so many peculiarities,
as noted above?  All of these phenomena (photometric fading by 2 mag over
2 centuries, ``veiling" of the optical spectra, intense IR emission, low-excitation
emission lines) are related to the rich circumstellar environment caused by the
star's very high mass-loss rate.  We have recently shown (Massey et al.\ 2005) that
in general the dust production rate of RSGs is proportional to the bolometric
luminosity of the star, but Danchi et al.\ (1994) in particular has emphasized that
the dust production is quite sporadic, with time scales on the order of a few decades.
Could the mass-loss rates of RSGs vary so significantly that VY CMa is simply
a normal RSG going through a short period of intense mass-loss that is normal?
We have called attention here to a newly noted peculiarity, namely that the star's
brightness seems to change by large amounts with little change in colors---even the
$V-K$ colors.  This is unlikely due to dust (due to the grayness), 
and suggests a luminosity change at nearly constant effective temperature. 
We note that since RSGs are fully convective, they do (by definition) lie on the edge
of the Hayashi forbidden zone.  If some instability caused the star to venture slightly
into this zone, we would expect the star to undergo a very short, unstable period.
We speculate that this might be responsible for variations in the star's luminosity and
in driving the dust production rate.
Possibly VY CMa, and the more extreme IR object NML Cyg, are examples of
normal RSGs that we have simply caught during an unusual time.

\acknowledgements

Our attention was originally called to this interesting star by John Monnier and
Roberta Humphreys.  We are also grateful for correspondence with George Herbig and George Wallerstein, whose seminal 
papers on this star made for interesting and enjoyable
reading.   We thank Knut Olsen for useful comments on the manuscript,
and for his continued
collaboration on the overall project of which this is a small part.
This paper made use of data products from the Two Micron All Sky Survey, which is a joint project of
the University of Massachusetts and the Infrared Processing and Analysis
Center/California Institute of Technology, funded by the National Aeronautics and 
Space Administration and the National Science Foundation.   We acknowledge with
thanks the variable star observations from the AAVSO International Database
contributed by observers worldwide and used in this research.  An anonymous
referee made useful comments on the paper, leading to improvements in the final version.

\clearpage

\begin{deluxetable}{l r r r r r r}
\tabletypesize{\scriptsize}
\tablewidth{0pc}
\tablenum{1}
\tablecolumns{7}
\tablecaption{\label{tab:results} Physical Properties VY CMa}
\tablehead{
\colhead{Method}
&\colhead{$T_{\rm eff}$ (K)}
&\colhead{$A_V$}
&\colhead{$\log g$[cgs]}
&\colhead{$M_V$}
&\colhead{$M_{\rm bol}$}
&\colhead{$R/R_\odot$}
}
\startdata                           
TiO (red/NIR bands)\tablenotemark{a} &$3650\pm25$     &3.20   &0.1   &$-5.6$ &$ -7.2$ & 605 \\
TiO (blue/yellow bands)                        &$3450\pm25$     &2.00  & 0.1   &$-4.4$ & $-6.9$ & 595 \\
V-K  &$>3475\pm35$     &3.20\tablenotemark{b}&  -0.2   &$-5.6$\tablenotemark{b}& $>-8.0$ & $<$955 \\
V-J   &$3705\pm90$&3.20\tablenotemark{b}     & 0.2   & $-5.6$\tablenotemark{b}& $-7.0$ & 545 \\
\enddata
\tablenotetext{a}{Preferred; see text.}
\tablenotetext{b}{Adopted.}
\end{deluxetable}

\clearpage

\begin{figure}
\epsscale{0.8}
\plotone{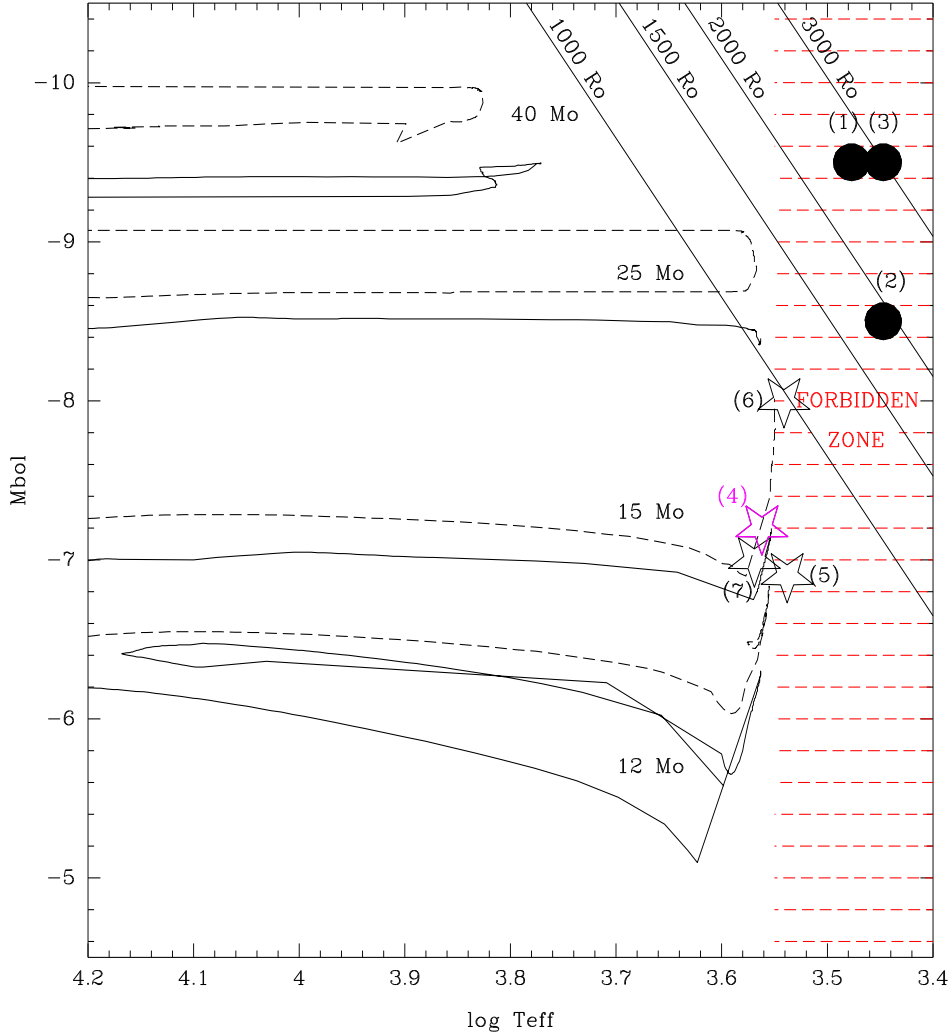}
\caption{\label{fig:HRD} Location of VY CMa in the H-R Diagram.
The three numbered filled circles correspond to the ``old" parameters of
VY CMa, where (1) is from Smith et al.\ (2001), (2) is from Monnier et al.\ (1999),
and (3) is from Le Sidaner \& Le Bertre (1996), where we have adjusted their
luminosity to our adopted 1.5~kpc distance.  In all three cases the effective temperature
has simply been assumed based upon the spectral type.  All three points lie well into
the ``forbidden zone" (hashed red area)
to the right of the Hayashi limit, and so are hydrostatically
unstable.  The five-starred points correspond to the results of this paper: 
(4)  is based on our solution fitting the NIR TiO bands;
(5) is based upon our solution fitting the blue TiO bands;
(6) is based upon based upon  the $(V-K)_0$ photometry (and which we take to be
a lower limit on the temperature, and an upper limit on the luminosity); and
(7) is based upon the $(V-J)_0$ colors.  We denote our preferred values (point 4) in purple.
The evolutionary tracks
are from Meynet \& Maeder (2005), with solid lines indicating no initial
rotation, and dashed lines indicating the tracks with initial rotations of 300 km s$^{-1}$.
The diagonal lines at upper right show lines of constant radii.}
\end{figure}

\clearpage

\begin{figure}
\epsscale{0.49}
\plotone{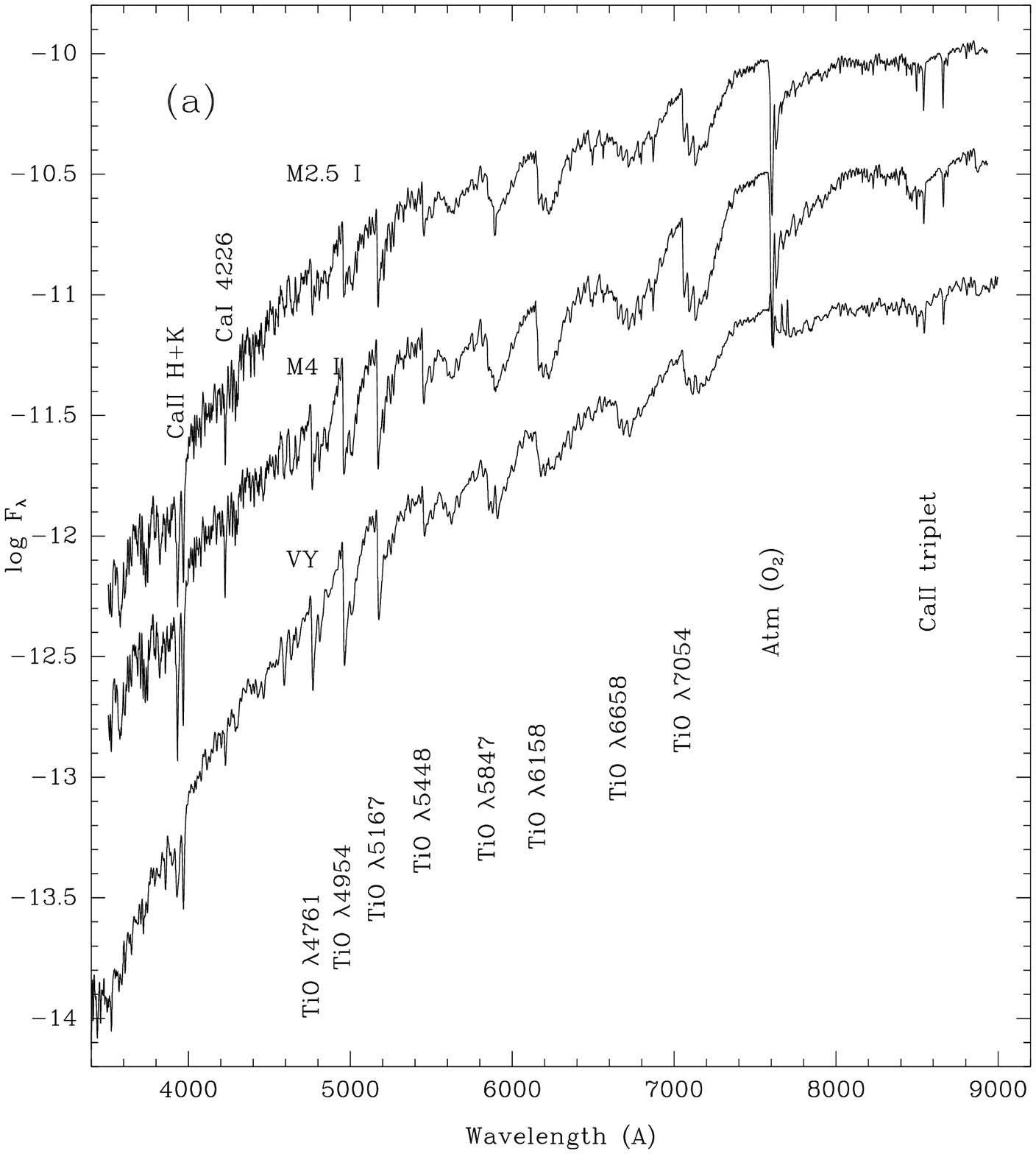}
\plotone{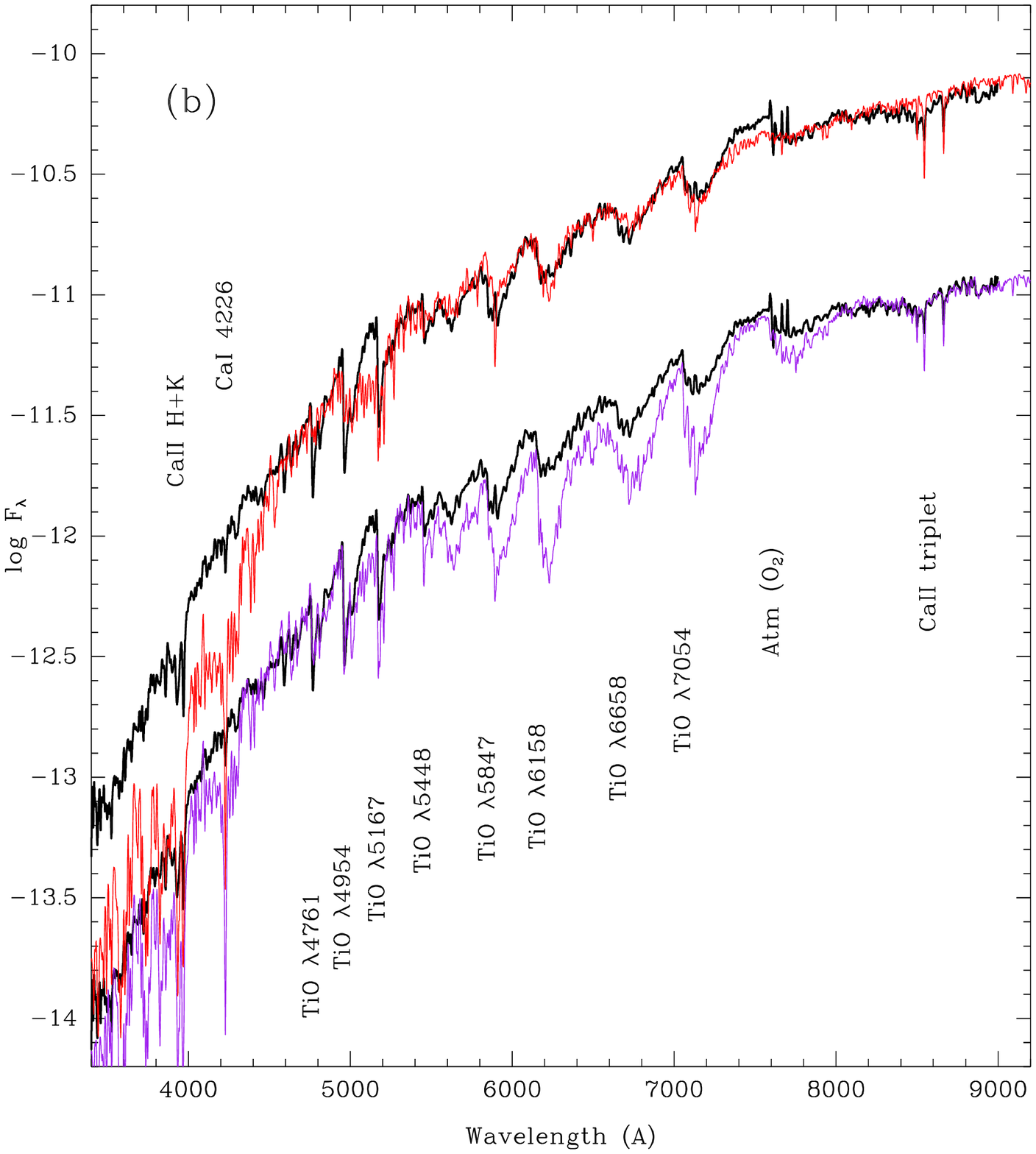}
\caption{\label{fig:fit} The spectrum of VY CMa. In (a) we compare
the spectrum of VY CMa (bottom spectrum) to that of the M2~I star HD 100930 (top
spectrum)
and the M4~I star HD 93420 (middle spectrum), where the data for the two
standards comes from Paper~I, and have not been corrected for telluric features.
The standards have been shifted vertically by arbitary amounts.
In (b) we show the model fits.
The red curve is our preferred model fit 
($T_{\rm eff}$=3650~K and $A_V=3.20$), which does
a good job reproducing the TiO bands in 
the red (i.e., TiO $\lambda \lambda$ 6158, 6658, 7054), 
but produces TiO bands too weak in the blue (i.e., TiO $\lambda \lambda$ 4761, 4954, 5167). The spectrum (and model fit) have been shifted vertically.
The lower spectrum shows the ``fit"  (purple)  with
a cooler model ($T_{\rm eff}$=3450) 
and less reddening ($A_V=2.00$). The TiO bands
in the blue are in better agreement, but the model produces TiO bands that are much too strong in the red.  As found
for other RSGs with significant circumstellar dust, 
there is significant extra flux in the observed stellar continuum in the near-UV, due, we believe,
to scattering by the dust (Massey et al.\ 2005).  The models also do not do a good job of
reproducing the atomic CaI and CaII features, as discussed in Paper~I.  
}
\end{figure}

\clearpage

\begin{figure}
\epsscale{0.8}
\plotone{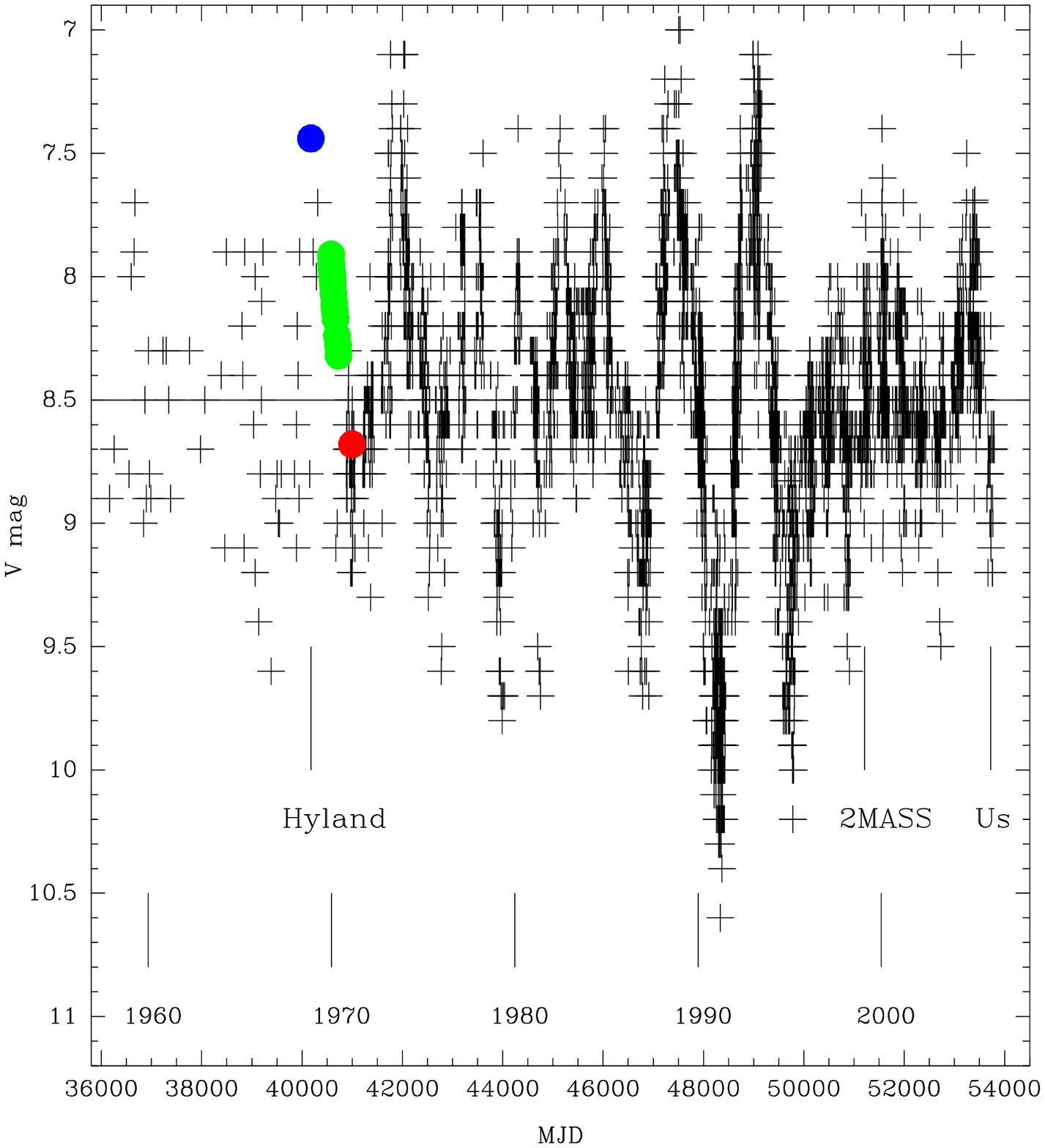}
\caption{\label{fig:aavso} Visual photometry of VY CMa.  The ``+" symbols are visual estimates
(typically $\pm0.2$~mag) from the AAVSO (Henden 2006).  
The colored filled circles are photoelectric $V$ measurements: the blue point
is that quoted by Hyland et al.\ (1969); the 
green points are from Cousins \& Lagerweij (1971), and the red point is from
that quoted by Wallerstein (1971).  These indicate good agreement with the visual data. We indicate the times of the IR photometry (Hyland et al.\ (1969) and 2MASS)
and of our spectrophotometry.
The solid line at $V=8.5$ shows the median value throughout the AAVSO data set.
}
\end{figure}


\begin{references}
\reference {} Bessell, M. S. PASP, 111, 1426
\reference {} Bessell, M. S., \& Brett, J. M. 1988, PASP, 100, 1134
\reference {} Bohm-Vitense, E.  1992, Introduction to Stellar Astrophysics, Volume 3 Stellar
Structure and Evolution (Cambridge: Cambridge Univ.\ Press), 134
\reference {} Cardelli, J., Clayton,G. D., \& Mathis, J. S. 1989, ApJ, 345, 245
\reference {} Carpenter, J. M. 2001, AJ, 121, 2851
\reference {} Cousins, A. W. J., \& Lagerweij, H. C. 1971, MNSSA, 30, 12
\reference {} Danchi, W. C., Bester, M., Degiacomi, C. G., Greenhill, L. J., \& Townes, C. H. 1994, AJ, 107, 1469
\reference {} Dyck, H. M., Lockwood, G. W., \& Capps, R. W. 1974, ApJ, 189, 89
\reference {} Eliasson, B., \& Bartlett, J. F. ApJ, 155, L79
\reference {} Ezer, D., \& Cameron, A. G. W. 1967, Can. J. Phys. 45, 3429
\reference {} Gustafsson, B., Bell, R. A., Eriksson, K., \& Nordlund, \AA\ 1975, A\&A, 42, 407
\reference {} Gustafsson, B., Edvardsson, B., Eriksson, K., Mizumo-Wiedner, M., Jorgensen,
U. G., \& Plez, B. 2003, in ASP Conf.\ ser.\ 288, Stellar Atmosphere
Modeling, ed. I. Hubeny, D., Mihalas, \& K. Werner (San Francisco: ASP), 331
\reference {} Henden, A. A. 2006,  Observations from the AAVSO International Database
\reference {} Henyey, L. G., Vardya, M. S., \& Bodenheimer, P. L. 1965, ApJ, 142, 841
\reference {} Herbig, G. H. 1970a, ApJ, 162, 557
\reference {} Herbig, G. H. 1970b, Evolution Stellaire Avant la Sequence 
Principale (Mem. Soc. Roy. Sci. Liege, 19), 13
\reference {} Herbig, G. H. 1972, ApJ, 172, 375
\reference {} Herbig, G. H. 1974, ApJ, 188, 533
\reference {} Humphreys, R. M. 1974, ApJ, 188, 75
\reference {} Humphreys, R. M., Davidson, K., Ruch, G., \& Wallerstein, G. 2005, AJ, 129, 492
\reference {} Humphreys, R. M., \& Lockwood, G. W. 1972, ApJ, L59
\reference {} Hyland, A. R., Becklin, E. E., Neugebauer, G., \& Wallerstein, G.
1969, ApJ, 158, 619
\reference {} Jaschek, C., \& Jaschek, M. 1990, The Classification of Stars
(Cambridge: Cambridge Univ.~ Press), 350
\reference {} Joy, A. H. 1942, ApJ, 96, 344
\reference {} Kippenhahn, R., \& Weigert, A. Stellar Structure and Evoution (Berlin: Springer-Verlag),
228
\reference {} Lada, C. J., \& Reid, M. J. 1978, ApJ, 219, 95
\reference {} Le Sidaner, P., \& Le Bertre, T. 1996, A\&A, 314, 896
\reference {} Levesque, E. M., Massey, P., Olsen, K. A. G., Plez, B.,
Josselin, E., Maeder, A., \& Meynet, G. 2005, ApJ, 628, 973 (Paper I)
\reference {} Levesque, E. M., Massey, P., Olsen, K. A. G., Plez, B., Meynet, G., \& Maeder, A. 2006,
ApJ, 645, in press (Paper II)
\reference {} Massey, P. Plez, B., Levesque, E. M., Olsen, K. A. G.,
Clayton, G. C., \& Josselin, E. 2005, ApJ, 634, 1286 
\reference {} Meynet, G., \& Maeder, A. 2005, A\&A, 429, 581
\reference {} Merrill, P. W., Deutsch, A. J., \& Keenan, P. C. 1962, ApJ, 136, 21
\reference {} Monnier, J. D., Tuthill, P. G., Lopez, B., Cruzalebes,
P., Danchi, W. C., \& Haniff, C. A. 1999, ApJ, 512, 351
\reference {} Monnier, J. D. et al.\ 2004, ApJ, 605, 436
\reference {} Plez, B. 2003, in ASP Conf.\ Ser. \ 298, {\it GAIA} Spectroscopy: Science
and Technology, ed. U. Munari (San Francisco: ASP), 189
\reference {} Plez, B., Brett, J. M, \& Nordlund, \AA\  1992, A\&A, 256, 551
\reference {} Richards, A. M. S., Yates, J. A., \& Cohen, R. J. 1998, MNRAS, 299, 319\reference {} Robinson, L. J. 1970, IBVS 465
\reference {} Robinson, L. J. 1971, IBVS 599
\reference {} Schlegel, D. J., Finkbeiner, D. P., \& Davis, M. 1998, ApJ, 500, 525
\reference {} Smith, N., Humphreys, R. M., Davidison, K., Gehrz,
R. D., Schuster, M. T., \& Krautter, J. 2001, AJ, 121, 1111
\reference {} Wallerstein, G. 1958 PASP, 70, 479
\reference {} Wallerstein, G. 1977 ApJ, 211, 170
\reference {} Wallerstein, G. 1978, Observatory, 98, 1026
\reference {} Wallerstein, G., \& Gonzalez, G. 2001, PASP, 113, 954
\reference {} Wilson, W. J., \& Barrett, A. H. 1968, Sci, 161, 778
\reference {} Worley, C. E. 1972, ApJ, 175 L93

\end{references}
\end{document}